\documentclass[twocolumn,seceq]{jpsj2}

\title{Gap Estimation by means of Hyperbolic Deformation}

\author
{Hiroshi {\sc Ueda}$^1$, Hiroki {\sc Nakano}$^2$,
Koichi {\sc Kusakabe}$^1$, and Tomotoshi {\sc Nishino}$^3$ }

\inst
{
$^{1)}$ Department of Material Engineering Science, 
Graduate School of Engineering Science, Osaka University, Osaka 560-8531\\
$^{2)}$ Graduate School of Material Science, University of Hyogo,
Kouto 3-2-1, Kamiori, Ako-gun, 678-1297\\
$^{3)}$ Department of Physics, Graduate School of Science, Kobe University, Kobe 657-8501
}

\recdate{\today}

\abst
{
We present a way of numerical gap estimation applicable for 
one-dimensional infinite uniform quantum systems. 
Using the density matrix renormalization group method 
for a non-uniform Hamiltonian, which has deformed interaction 
strength of $j$-th bond proportional to $\cosh \lambda j$, 
the uniform Hamiltonian is analyzed as a limit of $\lambda \rightarrow 0$. 
As a consequence of the deformation, an excited quasi-particle is 
weakly bounded around the center of the system, and kept away from the system boundary. 
Therefore, insensitivity of an estimated excitation gap of 
the deformed system to the boundary allows us to have the bulk excitation gap $\Delta(\lambda)$, 
and shift in $\Delta(\lambda)$ from $\Delta(0)$ is nearly linear in 
$\lambda$ when $\lambda \ll 1$. Efficiency of this estimation is demonstrated through 
application to the $S=1$ antiferromagnetic Heisenberg chain. 
Combining the above estimation and another one obtained from 
the technique of convergence acceleration for finite-size gaps 
estimated by numerical diagonalizations, 
we conclude that the Haldane gap is in $[0.41047905,~0.41047931]$. 
}
\kword{DMRG, renormalization group, excitation gap, antiferromagnetic Heisenberg chain, 
Haldane gap, exact-diagonalization method}

\begin{document}
\sloppy
\maketitle

\section{Introduction}
Analysis of elementary excitations has been one of the central concerns in 
the condensed matter physics. 
The ground state of an infinitely large 
quantum system, that has a finite excitation gap, 
is quite different from the gapless systems in its correlation properties. 
Precise estimation of the excitation gap is therefore important, 
particularly in numerical analysis of correlated systems. 

Because of limitation in computational resources, 
it is difficult to handle directly an infinite size system, 
but precise numerical analysis is possible on data for finite size systems. 
The method of finite size scaling (FSS) has been 
employed for the extrapolation of the data 
to the infinite size limit.~\cite{FSS1,FSS2} 

As an example of the gapped system, 
let us consider a spin-$S$ antiferromagnetic Heisenberg chain. 
When $S$ is an integer, the system has a nonzero excitation energy $\Delta$, 
which is known as the Haldane gap.~\cite{Haldane,Haldane2,NB} 
In estimation of this gap, finite size corrections should be 
subtracted properly from the numerical data. 
Reliability of such extrapolation procedure is 
partially dependent on the maximum 
of available system size that is handled by computation resources. 
In the case of the $S = 1$ chain, the maximum at present is 
around 24 by use of the Lanczos diagonalization~\cite{Lin,Golinelli2,Nakano}, 
but it becomes thousands by use of 
the density matrix renormalization group (DMRG) 
method~\cite{White1,White2,Peschel,Schollwoeck}. 

Appropriate choice of the boundary condition 
is an important procedure 
for the precise estimation of the excitation gap. 
In case of a one-dimensional system with open boundary conditions,
reflection at the system boundary occasionally gives a nontrivial
contribution to the kinetic energy of the excited quasi particle,
while the particle itinerates in the whole system.
Such a reflection effect can be reduced by means of a fine tuning of the 
boundary condition. In case of the $S = 1$ chain, an efficient way is to put 
an additional $S = 1 / 2$ spin at each end of the system, 
and to reduce the value of $J_{\rm end}^{~}$, 
the coupling constant between the $S = 1$ and the $S = 1 / 2$ spins 
at the boundary, 
compared with $J$, the exchange interaction inside the system.
The value $\Delta = 0.41050(2)$ was reported under the condition
$J_{\rm end}^{~} = 0.5088$.~\cite{Huse} 
In order to obtain a precise reference 
data for the following study in this article, 
we swept the value of $J_{\rm end}^{~}$, 
and obtained a slightly smaller value $\Delta = 0.41047944(27)$ when 
$J_{\rm end}^{~} = 0.50866$,~\cite{order} as shown in this paper. 
It should be noted that this kind of fine tuning 
at the system boundary is necessary for each system under study. 
For example, if the $S = 1$ chain contains uniaxial anisotropy, 
the most appropriate value of 
$J_{\rm end}^{~}$ is dependent on the anisotropy parameter. 

In this article we propose a way of erasing the boundary reflection effect, by 
weakly confining the excited quasiparticle around the center of the system. 
For this purpose, we introduce the so-called hyperbolic deformation to the 
one-dimensional quantum Hamiltonians, where interaction strength between
neighboring sites is proportional to $\cosh \lambda j$. Here, $j$ is
the lattice index running from $-\infty$ to $\infty$, 
and $\lambda$ is the deformation parameter.~\cite{Ueda} 
Although the interaction strength becomes position dependent, 
the ground state preserves 
a uniform property for any positive $\lambda$. 
For example, the expectation
value of the bond energy of the deformed Heisenberg chain 
is almost position independent. 
This uniform property in the ground state can be explained from 
the geometrical interpretation of the hyperbolic deformation. 

The effect of non-uniformity in the deformed Hamiltonian appears in the
elementary excitation. As we show in the following study on the deformed 
$S = 1$ Heisenberg chain, an excited quasiparticle
is weakly attracted to the center of the system, where the width of the 
bound state is proportional to $1 / \sqrt{\lambda}$. 
The corresponding excitation gap 
$\Delta(\lambda)$ is nearly linear in $\lambda$ when $\lambda \ll 1$. 
It is shown that the extrapolation of $\Delta(\lambda)$ 
to the limit $\lambda \rightarrow 0$ 
accurately gives the Haldane gap. 
The obtained value is compared with 
another value determined by 
the sequence interval squeeze (SIS) method.~\cite{Nakano} 
The precise procedure of the SIS method developed by Nakano and Terai
is explained in this article and another application of this method is given.
This technique with the exact numerical diagonalization 
gives bounds for the Haldane gap. 
We will conclude definitely that the upper bound of the Haldane gap 
is given by $\lim_{\lambda\rightarrow 0} \Delta(\lambda)$ much precisely 
than the SIS method. 
In this paper, 
the present best estimation of the lower bound given by 
the SIS method is also given. 

In the next section we explain the geometric background of the hyperbolic 
deformation. As an example, we consider a deformed tight-binding model, 
and its continuum limit. 
In \S 3, we show the distribution of the magnetic quasiparticle 
under the deformation, where the observed shallow bound state is in 
accordance with the tight-binding picture. In \S 4, we perform 
extrapolation $\lambda \rightarrow 0$ for the estimation of the 
excitation gap $\lim_{\lambda \rightarrow 0}\Delta( \lambda )$. 
As an independent estimate of the Haldane gap, 
we give the present best result by the SIS method in \S 5. 
Conclusions are summarized in the last section. 

\section{Hyperbolic Deformation}

Real- or imaginary-time evolution of a one-dimensional (1D)
quantum system is related
to a 2D classical system through so called the quantum-classical
correspondence.~\cite{Trotter,Suzuki,Baxter}
Our aim here is to generalize the correspondence for a general case
where the classical system is on curved 2D spaces.
Let us consider a hyperbolic plane, which is a 2D space
with constant negative curvature.
Suppose that there is a uniform classical field
on the hyperbolic plane, where the local action is position independent.
Then, how does the corresponding 1D quantum Hamiltonian look like?
We consider this problem for the case of imaginary-time evolution.

\begin{figure}[Htb]
\begin{center}
\includegraphics[width=50mm]{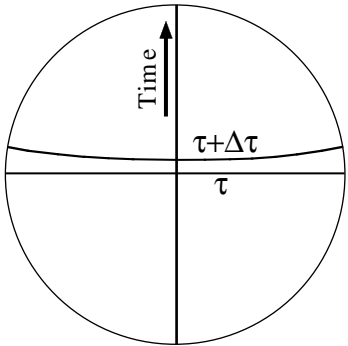}
\end{center}
\caption{\label{hyperbolic}
Imaginary-time axis and equal-time lines in the hyperbolic plane
drawn inside the Poincar\'e disc.
}
\label{fig1}
\end{figure}

Figure 1 shows the 2D hyperbolic space drawn inside the Poincar\'e
disc.
All the geodesics are represented by arcs, which are perpendicular to
the
border circle, including straight lines that pass through
the center of the disc.
Let us regard the vertical line as the imaginary-time axis.
Then all the geodesics that are
perpendicular to this imaginary-time axis can be regarded as equal-
time curves.
Suppose that the horizontal line corresponds to the coordinate $x$
of the quantum system,
and consider a quantum state $| \Psi( \tau ) \rangle$ on this
line.
If the classical action in the lower half of the hyperbolic plane is
uniform,
and if there is no symmetry breaking such as dimerization,
the state $| \Psi( \tau ) \rangle$ is also translationally
invariant.
This is because $| \Psi( \tau ) \rangle$
is given by imaginary time boost from $\tau = - \infty$,
which is mediated by the uniform action.

Let us consider an infinitesimal evolution
\begin{equation}
| \Psi( \tau + \Delta \tau ) \rangle = {\cal U}
[ \Delta \tau]  \,\, | \Psi( \tau ) \rangle
\label{time_evolution}
\end{equation}
from $\tau$ to $\tau + \Delta \tau$,
where ${\cal U}[ \Delta \tau]$ represents the
imaginary-time boost operation. Though
both $| \Psi( \tau ) \rangle$ and $| \Psi( \tau +
\Delta \tau ) \rangle$ are
translationally invariant,  ${\cal U}[ \Delta \tau]$ is not.
The fact can be seen geometrically by considering the distance between
two points $( x, \tau )$ and $( x, \tau + \Delta \tau )$
on the hyperbolic plane,
which is an increasing function of $| x |$.
The distance can be represented as
$( \cosh \nu x ) \, \Delta \tau$,~\cite{diverge}
where $\nu$ is a constant which is a function
of the scalar curvature of the hyperbolic plane. If it is possible to
represent ${\cal U}[ \Delta \tau ]$ in the exponential form
\begin{equation}
{\cal U}[ \Delta \tau ]
= \exp\left(-\int \hat{h}(x) ( \cosh \nu x )
\, \Delta \tau d x \right)
= \exp( - \Delta \tau { H} ) \, ,
\label{time_boost_ope}
\end{equation}
the corresponding Hamiltonian $H$ is also position dependent.
In this case, $H$ is written by an integral of a local operator $
{\hat h}( x )$,
and the position dependence is explicitly written as
\begin{equation}
{ H} = \int ( \cosh \nu x ) \, \hat{h}( x )
\, d x \, .
\label{continuous_ham}
\end{equation}
This is an example of the {\it hyperbolic deformation} of quantum
Hamiltonian
in the continuous 1D space.
If ${\hat h}( x )$ contains derivatives with respect to $x$, 
the form of $H$ becomes a complicated one.
So, let us introduce explicit construction of the hyperbolic deformation
starting from a microscopic Hamiltonian.

Consider a way of introducing the hyperbolic deformation to the
lattice systems.
We introduce lattice points at $x = a j$, where $a$ is the lattice
constant and $j$ is the lattice index, which runs in a finite
range from $-R/a$ to $R/a$.
The constant $R$ we have introduced satisfies $R/a \gg 1$, and
specifies the size of the system, which ensures a numerical cutoff.
Then we have a relation
\begin{equation}
\cosh \nu x = \cosh \nu a j = \cosh \lambda j \, ,
\label{lattice_sytem}
\end{equation}
where $\lambda = \nu a$ is the deformation parameter that we have
used.
A discrete analogue of ${ H}$ in eq.~(\ref{continuous_ham}) is
then given by
the following lattice Hamiltonian
\begin{equation}
H( \lambda )
=
\sum_j^{~} \cosh \lambda j   \,\, h_{j, j+1}^{~}
+
\sum_j^{~} \cosh \lambda \bigl( j - {\textstyle
\frac{1}{2}} \bigr)  \,\, g_j^{~} \, ,
\label{lattice_ham}
\end{equation}
where $h_{j, j+1}^{~}$ represents the neighboring interaction,
$g_j^{~}$ the on-site ones. Another possible choice of the discrete
Hamiltonian is
\begin{eqnarray}
&&H( \lambda )
=
\sum_j^{~} \cosh \lambda j   \,\, h_{j, j+1}^{~}
\nonumber\\
&& ~~~~~ + \frac{1}{2}  \sum_j^{~} \biggl[
\cosh \lambda j  + \cosh \lambda ( j - 1 )
\biggr] \,\, g_j^{~} \\
&&= \sum_j^{~} \cosh \lambda j   \,\, h_{j, j
+1}^{~} + \cosh \frac{\lambda}{2}
\sum_j^{~} \cosh \lambda \bigl( j - {\textstyle
\frac{1}{2}} \bigr)  \,\, g_j^{~} \, ,
\nonumber
\label{lattice_ham2}
\end{eqnarray}
where the coefficient of the on-site term is different from
eq.~(\ref{lattice_ham}).
Since we chiefly investigate small $\lambda$ region, this difference
is
not conspicuous.~\cite{long}
We therefore choose $H( \lambda )$ in the form of
eq.~(\ref{lattice_ham}) in the
following. It should be noted that 
the Hamiltonian $H( \lambda = 0 )$ is translationally invariant.

As an example of the 1D lattice systems, let us consider a non- 
interacting
tight-binding model. The deformed Hamiltonian is written as
\begin{eqnarray}
H_{\rm TB}^{~}( \lambda ) &=& - t
\sum_j^{~} \cosh \lambda j  \left(
\, c_j^\dagger c_{j+1}^{~} +
\, c_{j+1}^\dagger c_j^{~} \right) \nonumber\\
&& - \mu \sum_j^{~} \cosh \lambda \bigl( j -
{\textstyle \frac{1}{2}} \bigr)
\, c_j^\dagger c_j^{~} \, ,
\label{tight_binding}
\end{eqnarray}
where $t$ represents the hopping amplitude, and $\mu$ the chemical
potential. The operators, $c_j^\dagger$ and $c_j^{~}$,
appearing in eq.~(\ref{tight_binding})
are fermion creation and annihilation operators.~\cite{commute}
Since there is no interaction,
all the eigenstates can be constructed from one-particle
wave functions $\Psi_j^{~} = \langle j | \Psi \rangle$,
where $| \Psi \rangle$ is a 1-particle eigenstate and $\langle
j |$ is defined
as $\langle 0 | c_j^{~}$. The wave function $\Psi_j^{~}$
of the stationary state satisfies the
Schr\"odinger equation
\begin{eqnarray}
E \, \Psi_j^{~} &=&
- t \cosh \lambda j \, \Psi_{j+1}^{~}
- t \cosh \lambda \bigl( j - 1 \bigr) \,
\Psi_{j-1}^{~} \nonumber\\
&~&
- \mu \cosh \lambda \bigl( j - {\textstyle \frac{1}
{2}} \bigr) \, \Psi_j^{~} \, .
\label{sch_eq}
\end{eqnarray}
The one-particle ground state energy $E_0^{~}$ becomes zero when
$- \mu / t = 2 \cosh (\lambda / 2)$,
and the corresponding wave function $\Psi_j^{~}$ becomes a
constant of $j$.
If $\mu$ is smaller than $- 2 t \cosh (\lambda / 2)$, the
ground state
wave function is bounded around the origin $j = 0$. This kind of bound
state is also observed for one particle excitation in many body problem,
as we will see in the next section.

Let us check the continuum limit of eq.~(\ref{sch_eq}).
Substituting the relations $x = a j$, $\lambda = a \nu$ and
the correspondence
\begin{equation}
\Psi_j^{~} = \Psi( a j ) = \Psi( x )
\label{correspondence}
\end{equation}
to eq.~(\ref{sch_eq}), we obtain the relation
\begin{eqnarray}
&& E \,  \Psi( x ) = \nonumber\\
&~&
- t  \cosh \nu \bigl( x - \frac{a}{2} \bigr)
\cosh \nu \frac{a}{2}
\bigl[ \Psi( x + a ) + \Psi( x - a ) \bigr]
\nonumber\\
&~&
- t  \sinh \nu \bigl( x - \frac{a}{2} \bigr)
\sinh \nu \frac{a}{2}
\bigl[ \Psi( x + a ) - \Psi( x - a ) \bigr]
\nonumber\\
&~&
- \mu \cosh \nu \bigl( x - \frac{a}{2} \bigr)
\Psi( x )
\label{sch_eq2}
\end{eqnarray}
after some algebra. Expressing the hopping amplitude as
$t = \hbar^2 / (2 m a^2_{~})$,
chemical potential as $\mu = - U - 2 t$, and
taking the limit $a \rightarrow 0$, we obtain a differential equation
\begin{equation}
E \, \Psi( x ) = \left[
- \frac{\hbar^2}{2m}  \frac{\partial}{\partial x}
\cosh \nu x \, \frac{\partial}{\partial x}
+ U \cosh \nu x \right] \Psi( x ) \, .
\label{sch_eq3}
\end{equation}
The first term in the parenthesis of the r.h.s. is the deformed
kinetic energy,
and the second term is a kind of trapping potential when $U > 0$.

If we consider the imaginary-time dependence of the wave function, the
Lagrangian which draws eq.~(2.11) from the stationary condition is
given by
\begin{eqnarray}
\lefteqn{{\cal L}(\Psi^*,\partial_{\tt t}\Psi^*,
\partial_x\Psi^*,\Psi,\partial_{\tt t}\Psi,
\partial_x\Psi)}\nonumber \\
& = &
\Psi^*_{~} \frac{\partial}
{\partial {\tt t}} \Psi +
\cosh \nu x \left[
\frac{\hbar^2}{2m} \frac{\partial \Psi^*_{~}}
{\partial x} \frac{\partial \Psi}{\partial x} +
U \Psi^*_{~} \Psi
\right]
\nonumber \\
\label{lagrangian}
\end{eqnarray}
for $\Psi( x, {\tt t} )$,
where we have introduced the letter ${\tt t}$ for the imaginary-time  
variable,
and where we have used the unit that satisfies $\hbar = 1$.
Note that the time-like variable $\tau$ in eq.~(2.1) is related
to ${\tt t}$ by
the relation $( \cosh \nu x ) \, d {\tt t} =
d \tau$,
and in the $x$-$\tau$ plane
the Lagrangian can be represented as
\begin{eqnarray}
\lefteqn{{\cal L}'(\Psi^*,\partial_{\tau}\Psi^*,
\partial_x\Psi^*,\Psi,\partial_{\tau}\Psi,
\partial_x\Psi)}\nonumber \\
& = &
\cosh \nu x \left[
\Psi^*_{~} \frac{\partial}{\partial \tau} \Psi +
\frac{\hbar^2}{2m} \frac{\partial \Psi^*_{~}}
{\partial x} \frac{\partial \Psi}{\partial x} +
U \Psi^*_{~} \Psi
\right]
\, ,
\nonumber \\
\label{lagrangian2}
\end{eqnarray}
for $\Psi( x, \tau )$.
The action of the system is given by
\begin{eqnarray}
S&=&\int {\cal L}'(\Psi^*,\partial_{\tau}\Psi^*,
\partial_x\Psi^*,\Psi,\partial_{\tau}\Psi,
\partial_x\Psi) \, d\tau dx \nonumber \\
&=&\int
\left[\Psi^*_{~} \frac{\partial}{\partial
\tau} \Psi + \hat{h}(x) \right]
(\cosh \nu x) \, d\tau dx.
\end{eqnarray}
This action is actually obtained by
identifying $\Psi(x, \tau)$ as a field operator and
deriving the path-integral formalism starting from eq.~(2.1)
with a local Hamiltonian
\begin{equation}
\hat{h}(x)=
\frac{\hbar^2}{2m} \frac{\partial \Psi^*_{~}}
{\partial x}
\frac{\partial \Psi}{\partial x} + U \Psi^*_{~}
\Psi.
\end{equation}
We note that a local deformation
of the measure in the action
gives the hyperbolic deformation.

\section{Excitation of the $S = 1$ Heisenberg Chain}
\label{result1}

We consider the $S = 1$ antiferromagnetic Heisenberg chain as an example of the 
1D many body systems. The system has finite magnetic excitation energy,
which is known as the Haldane gap.~\cite{Haldane,Haldane2,NB} In numerical
analyses to obtain the gap of an open-boundary system,
there is a custom to put $S = 1/2$ spins at 
both ends of the system, in order to avoid the quasi 
degeneracy in the low-energy states.~\cite{Huse,Kennedy}  
The Hamiltonian of the open-boundary $S = 1$ chain is represented as
\begin{eqnarray}
H(0) &=& J  \sum_{j = - N + 1}^{N - 1} {\bf S}_j^{~} \cdot {\bf S}_{j+1}^{~} 
\label{Heisenberg1}
\\
&+& J_{\rm end}^{~} \,  \left( 
{\bf s}_{\rm L}^{~} \cdot {\bf S}_{- N + 1}^{~} + 
{\bf S}_{N}^{~} \cdot {\bf s}_{\rm R}^{~} \right) \, , 
\nonumber
\end{eqnarray}
which includes $M=2N$ numbers of $S = 1$ spins from ${\bf S}_{-N+1}^{~}$ to 
${\bf S}_N^{~}$, and the boundary $S = 1 / 2$ spins ${\bf s}_{\rm L}^{~}$ and
${\bf s}_{\rm R}^{~}$. Thus there are $2N + 2$ spins in total. We count the 
number of $S = 1$ spins $M$ as the size of the system. 
The parameter $J > 0$ represents the 
antiferromagnetic exchange coupling between neighboring $S = 1$ spins 
${\bf S}_j^{~}$ and ${\bf S}_{j+1}^{~}$, 
and $J_{\rm end}^{~} > 0$ is the coupling at the boundary between 
${\bf s}_{\rm L}^{~}$ and ${\bf S}_{- N + 1}^{~}$ and also between
${\bf S}_{N}^{~}$ and ${\bf s}_{\rm R}^{~}$.
Throughout this article we take $J$ as the unit of the energy, 
and use the parameterization 
$J_{\rm end}^{~} = J = 1$ unless the value of $J_{\rm end}^{~}$ is specified.

We introduce the hyperbolic deformation for this system. 
The deformed Hamiltonian is represented as
\begin{eqnarray}
&& H( \lambda ) = J \!\!\! \sum_{j = - N + 1}^{N - 1} 
\cosh \lambda j   \,\,  {\bf S}_j^{~} \cdot {\bf S}_{j+1}^{~} 
\label{even}
\\
&& ~~~~ + J_{\rm end}^{~} \, \cosh \lambda N \left( 
{\bf s}_{\rm L}^{~} \cdot {\bf S}_{- N + 1}^{~} + 
{\bf S}_{N}^{~} \cdot {\bf s}_{\rm R}^{~} \right) \, .
\nonumber
\end{eqnarray}
When $\lambda=0$, eq. (\ref{even}) becomes eq. (\ref{Heisenberg1}).
Occasionally it is convenient to treat a system that contains 
odd number of spins,
so that one of the $S = 1$ spin is just at the center of the system.
In order to satisfy the condition, we introduce another type of the 
deformed system described by the Hamiltonian
\begin{eqnarray}
&&H( \lambda ) = J \, \sum_{j = - N + 1}^{N - 2} 
\cosh \lambda \bigl( j - {\textstyle \frac{1}{2}} \bigr)  \,  {\bf S}_j^{~} \cdot {\bf S}_{j+1}^{~} 
\label{odd}
\\
&&+ J_{\rm end}^{~} \cosh \lambda \bigl( N - {\textstyle \frac{1}{2}} \bigr) \left( 
{\bf s}_{\rm L}^{~} \cdot {\bf S}_{- N + 1}^{~} + 
{\bf S}_{N - 1}^{~} \cdot {\bf s}_{\rm R}^{~} \right)  \, , \nonumber
\end{eqnarray}
where there are $2N + 1$ spins in total. In this case the system size, which
is the number of $S = 1$ spins, is $M=2N - 1$.

\subsection{Gap estimation for the undeformed system}
\label{undeformed}

\begin{table}
\caption{\label{white_huse_gap_list}
Calculated excitation energy $\Delta_{M}^{~}$ 
for each system with the size $M$, which is the number of $S = 1$ spins. 
The integer $m$ is the number of states kept. 
The interaction strength at the system boundary is 
$J_{\rm end} = 0.5088$ (the upper series from $M=2N=100$ to 160) 
or $J_{\rm end} = 0.50866$ (the lower series).
Truncation errors $1 - P_{\rm GS}(m)$ and $1 - P_{\rm EX}(m)$ are 
also shown for the ground and the 1st excited states, respectively. 
(See text.)
}
\begin{center}
\begin{tabular}{ccccc}
\hline \hline
 $M$    & $m$ &  $\Delta_{M}^{~}$ & $1-P_{\rm GS}(m)$ & $1-P_{\rm EX}(m)$\\
\hline
\hline
100 & 120 &    0.4104951946 & 5.196E-12 & 7.724E-10\\
    & 140 &    0.4104949559 & 1.445E-12 & 3.009E-10\\
    & 160 &    0.4104948683 & 3.579E-13 & 1.556E-10\\
    & 180 &    0.4104948265 & 4.170E-14 & 6.870E-11\\
\hline
120 & 120 &    0.4104929134 & 5.194E-12 & 8.192E-10\\
    & 140 &    0.4104926043 & 1.446E-12 & 3.237E-10\\
    & 160 &    0.4104924896 & 3.769E-13 & 1.657E-10\\
    & 180 &    0.4104924345 & 4.125E-14 & 7.301E-11\\
\hline
140 & 120 &    0.4104912749 & 5.197E-12 & 8.539E-10\\
    & 140 &    0.4104908951 & 1.445E-12 & 3.300E-10\\
    & 160 &    0.4104907533 & 3.678E-13 & 1.732E-10\\
    & 180 &    0.4104906847 & 4.270E-14 & 7.615E-11\\
\hline
160 & 120 &    0.4104900535 & 5.193E-12 & 8.770E-10\\
    & 140 &    0.4104896021 & 1.443E-12 & 3.391E-10\\
    & 160 &    0.4104894323 & 3.695E-13 & 1.799E-10\\
    & 180 &    0.4104893502 & 5.500E-14 & 7.854E-11\\
\hline
\hline
100 & 160  &    0.4104803729 & 3.657E-13 & 1.562E-10\\
    & 180  &    0.4104803310 & 3.373E-13 & 6.875E-11\\
\hline
120 & 160  &    0.4104802346 & 3.692E-13 & 1.663E-10\\
    & 180  &    0.4104801793 & 4.313E-14 & 7.306E-11\\
\hline
140 & 160  &    0.4104801400 & 3.904E-13 & 1.738E-10\\
    & 180  &    0.4104800712 & 4.351E-14 & 7.620E-11\\
\hline
160 & 160  &    0.4104800736 & 3.690E-13 & 1.798E-10\\
    & 180  &    0.4104799915 & 6.541E-14 & 7.858E-11\\
\hline
\hline
\end{tabular}
\end{center}
\end{table}

We first estimate the value of the Haldane gap $\Delta$ 
for undeformed systems 
$\lambda = 0$, in order to get reference data for the later study under 
deformation $\lambda > 0$.
The excitation energy from the ground state is calculated by the DMRG 
method,~\cite{White1,White2,Peschel,Schollwoeck} as a function of $J_{\rm end}^{~}$,
the system size $M=2N$ (or $2N-1$), and the number of states kept $m$, 
which is increased up to 180. 
From the various values of $J_{\rm end}^{~}$ for which 
we have performed calculations, we show the data for two typical values 
$J_{\rm end}^{~} = 0.5088$ and $J_{\rm end}^{~} = 0.50866$ in Table \ref{white_huse_gap_list}.
The former value is used in a literature~~\cite{Huse} 
and the latter is an optimized one in this work.

The data for $J_{\rm end}^{~} = 0.5088$ in Table \ref{white_huse_gap_list} 
are calculated under the same conditions as them~\cite{Huse} except for 
a readjustment of the energy origin. 
For precise determination of the lowest eigenvalue, 
we shift the origin of the
energy so that the ground-state energy becomes nearly zero.
This energy shift is realized by the following process. 
First we obtain the ground state $| \Psi_0^{~} \rangle$
diagonalizing the Hamiltonian in eq.~(\ref{even}) or eq.~(\ref{odd}), 
and calculate the nearest neighbor correlation function 
$w_{i,i+1}^{~} = \langle \Psi_0^{~} | {\bf S}_i^{~} \cdot {\bf S}_{i+1}^{~} 
| \Psi_0^{~} \rangle$. We then replace the neighboring interaction 
${\bf S}_i^{~} \cdot {\bf S}_{i+1}^{~}$ in the Hamiltonian by 
${\bf S}_i^{~} \cdot {\bf S}_{i+1}^{~} - w_{i,i+1}^{~} \hat{I}$ 
with an identity operator $\hat{I}$, 
and perform the same subtraction also for the boundary terms where 
${\bf s}_{\rm L}^{~}$ and ${\bf s}_{\rm R}^{~}$ are involved.
This subtraction can be performed successively when one constructs the
renormalized Hamiltonians $H_{\rm L}^{~}$ and $H_{\rm R}^{~}$ for the left
and the right block of the system during the finite-size sweeping process.
The ground-state energy of the shifted Hamiltonian thus obtained is nearly zero.
The total amount of the energy shift can be obtained from
$w_{i,i+1}^{~}$. It should be noted that the above energy shift process
is important for the large-scale system, where the ground-state energy becomes
a big number. In the same manner, we have to use the shifted Hamiltonian 
when we consider the hyperbolically deformed system with $\lambda > 0$, 
where the absolute value 
of the ground-state energy increases exponentially with the system size.

Truncation errors introduced to the ground state $1~-~P_{\rm GS}(m)$,
which are listed in  Table \ref{white_huse_gap_list}, are calculated  
by the following standard procedures in DMRG calculation.
After sufficient numbers of finite size sweeping, we obtain the
optimized variational ground state
\begin{equation}
| \tilde{\Psi} \rangle = \sum_{\xi_l^{~} S_0^Z S_1^Z  
\xi_r^{~}}
\tilde{\Psi}_{\xi_l^{~} S_0^Z S_1^Z \xi_r^{~}}
| \xi_l^{~} \rangle | S_0^Z \rangle | S_1^Z \rangle |  
\xi_r^{~} \rangle \, ,
\end{equation}
where $|\xi_l\rangle$ and $|\xi_r\rangle$ represent  
relevant block spin
state for the left and the right blocks, respectively, that take at  
most $m$
numbers of states.
Creating the reduced density matrix for the left half of the system
\begin{equation}
\rho_{\xi_l^{~} S_0^Z; \, \xi_l' {S_0^Z}'} =
\sum_{S_1^Z  \xi_r^{~}}
\tilde{\Psi}^*_{\xi_l^{~} S_0^Z S_1^Z \xi_r^{~}}
\tilde{\Psi}_{\xi_l' {S_0^Z}' S_1^Z \xi_r^{~}} \, ,
\end{equation}
and diagonalizing it to obtain eigen values $w_\alpha$, where
we assume the ascending order for  $w_\alpha$.
The truncation error is then calculated as
\begin{equation}
1 - P_{\rm GS}(m) = 1 - \sum_{\alpha=1}^{m} w_\alpha~.
\end{equation}
The truncation error for the lowest excited state $1~-~P_{\rm EX}(m) 
$ is also
calculated in the same manner, using the optimized ground-state in the
subspace where the total number of $S^{Z}$ is equal to 1.


\begin{figure}[Htb]
\begin{center}
\includegraphics[width=85mm]{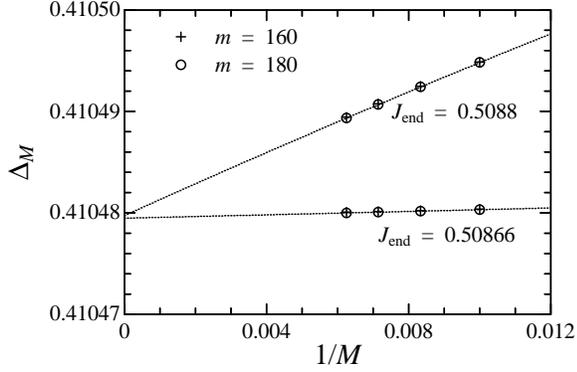}
\end{center}
\caption{System size dependence of the excitation gap $\Delta_{M}^{~}$ 
listed in Table \ref{white_huse_gap_list}. 
Only data for $m=160$ and $m=180$ is given.}
\label{white_huse_gap}
\end{figure}

Figure~\ref{white_huse_gap} shows the system size dependence of 
the excitation gap, which 
is given as a list in Table \ref{white_huse_gap_list}. For both cases, 
$J_{\rm end}^{~} = 0.5088$ and $J_{\rm end}^{~} = 0.50866$, the
gap is nearly linear in $1/M=1/(2N)$. 
From the numerical data under the condition $m = 160$, 
we obtain $\Delta = 0.41047970(1)$ when $J_{\rm end}^{~} = 0.5088$ and
$\Delta = 0.41047968(2)$ when $J_{\rm end}^{~} = 0.50866$ by 
use of the fitting with respect to second order polynomials, where
the numbers shown in the parenthesis are the mean-square fitting error.
These two values of the gap seem to be consistent, however, the 
gap is sensitive to the value of $m$.
From the data when we keep $m = 180$ states, we obtain
$\Delta = 0.41047947(1)$ when $J_{\rm end}^{~} = 0.5088$
and $\Delta = 0.41047944(1)$ when $J_{\rm end}^{~} = 0.50866$.
So far as we have calculated, the estimated value of $\Delta$ is always 
a decreasing function of $m$. 
This tendency is also discussed in the last section. 
Indeed, when $J_{\rm end}^{~} = 0.5088$, 
the values are 
$\Delta = 0.41048145(9)$ for $m = 120$, 
$\Delta = 0.41048019(4)$ for $m = 140$, 
$\Delta = 0.41047970(1)$ for $m = 160$, 
and 
$\Delta = 0.41047947(1)$ for $m = 180$. 
In this parameter range of $m$,  
change in $\Delta$ as a function of $m$ is bigger than 
the error in the last digit of the above estimation. 
Thus the lowest value obtained so far can be 
regarded as the upper bound for $\Delta$. 
We therefore use the smallest value $\Delta = 0.41047944$ 
when $m = 180$ and 
$J_{\rm end}^{~} = 0.50866$ as the better estimation 
of the upper bound of $\Delta$ than the estimation of 
$m = 180$ and $J_{\rm end}^{~} = 0.5088$. 

Note that the condition $J_{\rm end}^{~} = 0.50866$ is valid only for 
the Hamiltonians shown in Eqs.~(\ref{even}) and (\ref{odd}). 
If we need the excitation gap for a variety 
of $S = 1$ spin chains, 
which contain anisotropy and biquadratic terms, 
we have to find another appropriate value of $J_{\rm end}^{~}$ for each cases.

\subsection{Bounded excitation when $\lambda > 0$}

\begin{figure}[Htb]
\begin{center}
\includegraphics[width=80mm]{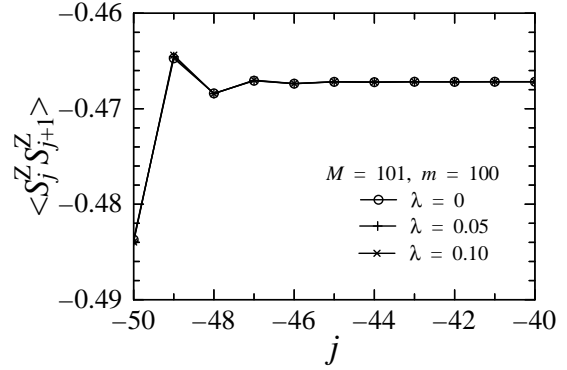}
\end{center}
\caption{
Nearest neighbor spin correlation function 
$\langle {S}_j^{\rm Z} {S}_{j+1}^{\rm Z} \rangle$ 
of the singlet ground state when 
the system size is $M=2N - 1 = 101$. 
 }
\label{nearest-neighbor} 
\end{figure}

Let us observe the ground state and the elementary excitation of the deformed 
chain, which is described by the Hamiltonian in eq.~(\ref{even}) 
or eq.~(\ref{odd}) with $\lambda > 0$. 
Figure~\ref{nearest-neighbor} shows the nearest-neighbor 
spin correlation functions between $S = 1$ spins 
$\langle {S}_j^{\rm Z} {S}_{j+1}^{\rm Z} \rangle$ 
of the singlet ground state when the system size is $M=2N -1 = 101$. 
In Fig.~\ref{nearest-neighbor}, we look only at the boundary of the system, 
where local fluctuation of this correlation function is prominent. 
It is known that the
hyperbolic deformation has an effect of decreasing the correlation
length $\xi$.~\cite{Ueda} 
In the $S=1$ antiferromagnetic 
Heisenberg spin chain, $\xi$ is of the order of unity 
already at $\lambda = 0$, and thus the effect of hyperbolic deformation 
is not conspicuous in long-range correlation functions, 
as long as the ground state is concerned. 
As displayed in Fig.~\ref{nearest-neighbor}, the short-range correlation 
function is also not affected by the deformation 
in this parameter range of $\lambda$. 
Although the interaction strength is position dependent, 
the spin correlation function 
is almost uniform inside the system. 
Thus we may say that the vacuum of the quasiparticle excitation is kept 
fixed in its internal structure against the deformation. 
This behavior is favorable to have a good convergence 
in the energy gap of the elementary excitation. 

\begin{figure}[Htb]
\begin{center}
\includegraphics[width=80mm]{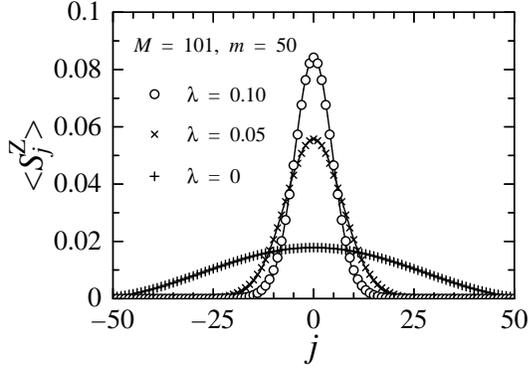}
\end{center}
\caption{
Spin polarization $\langle {S}_j^{\rm Z} \rangle$ of the first excited state.}
\label{spin_polarization} 
\end{figure}

Next, 
we observe the magnetic excitation. Figure~\ref{spin_polarization} 
shows the Z-component of the local spin polarization 
$\langle S_j^{\rm Z} \rangle$ calculated for the first
excited state, which is the 
lowest-energy state in the subspace 
where the total $S^Z_{~}$ of the system is unity.
The polarization  $\langle S_j^{\rm Z} \rangle$ is positive everywhere, 
unless one chooses an extremely large $\lambda$. Thus it 
is possible to regard $\langle S_j^{\rm Z} \rangle$ 
as the distribution probability of the
excited magnetic quasiparticle. 
The quasiparticle is bounded around the center of the system 
when $\lambda = 0.05$ and $0.10$, 
in contrast to the unbounded  case when $\lambda = 0$. 

\begin{figure}[Htb]
\begin{center}
\includegraphics[width=80mm]{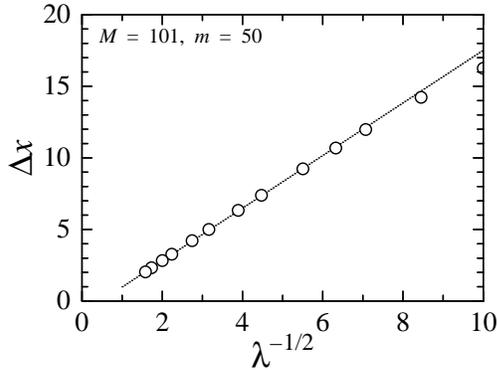}
\end{center}
\caption{
The distribution width $\Delta x$ of the quasiparticle with respect to $1 / \sqrt{\lambda}$.
}
\label{delta_x} 
\end{figure}

The observed quasiparticle distribution in Fig.~\ref{spin_polarization} 
is close 
to the Gaussian distribution around the origin $j = 0$, when $\lambda = 0.05$ 
or 0.10. 
In order to quantify the distribution width, we introduce 
\begin{equation}
\Delta x = \sum_{j = -N+1}^{N-1}  \sqrt{
j^2_{~} \, \langle S_j^{\rm Z} \rangle }
\label{distribution_width}
\end{equation}
for the cases when the system size is odd.
Figure~\ref{delta_x} shows the value of $\Delta x$ calculated for 
the $101$-site system. 
The dotted line is the fitting for those $\Delta x$ in  the 
range $\lambda^{-1/2}_{~} \leq 4$, equivalently $\lambda \geq 0.04$.
The distribution width $\Delta x$ is proportional to $1 / \sqrt{\lambda}$ 
in this parameter region of $\lambda$, 
where $\Delta x$ is almost independent of the system size. 
We have confirmed that
the relation $\Delta x \propto 1 / \sqrt{\lambda}$ holds for 
$1/\sqrt{\lambda} \leq 20$, equivalently $\lambda \geq 0.0025$, 
when the system size is $M=2N = 1000$.
If $\lambda$ becomes too small for a fixed system size, 
$\Delta x$ deviates from 
the fitting line due to the finite size effect. 
The deviation of $\Delta x$ from the $1 / \sqrt{\lambda}$ 
behavior suggests 
breaking the confinement of the quasiparticle. 

To speak qualitatively, the observed 
$\lambda$-dependence of $\Delta x$ 
is consistent with the 
effective one-particle potential 
\begin{equation}
U( j ) = J \, \cosh\lambda j  \, 
\sim \,  J + \frac{J}{2} \, ( \lambda j )^2_{~} 
\label{one-particle_potential}
\end{equation}
in the neighborhood of the origin $j = 0$. 
Note that a large system size $M$ ensures 
existence of a finite central region of the chain 
satisfying $\lambda j \ll 1$ for small but finite $\lambda$. 
Thus in a final simulation, we need to have an enough size $M\sim 1000$, 
which is tractable by the DMRG method at present. 

\begin{figure}[Htb]
\begin{center}
\includegraphics[width=80mm]{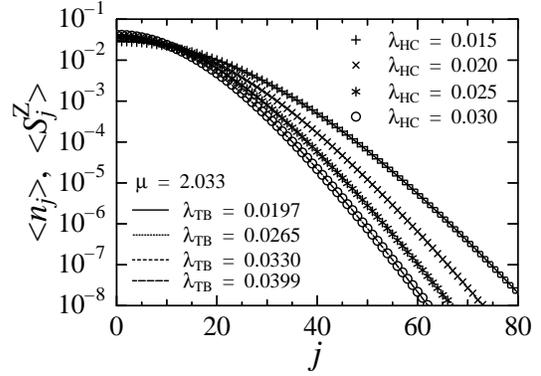}
\end{center}
\caption{ 
Comparison between $\langle n_j^{~} \rangle$ of the
tight-binding model and $\langle S_j^Z \rangle$ of the Heisenberg chain.}
\label{compair_hc_tb} 
\end{figure}

\begin{figure}[Htb]
\begin{center}
\includegraphics[width=80mm]{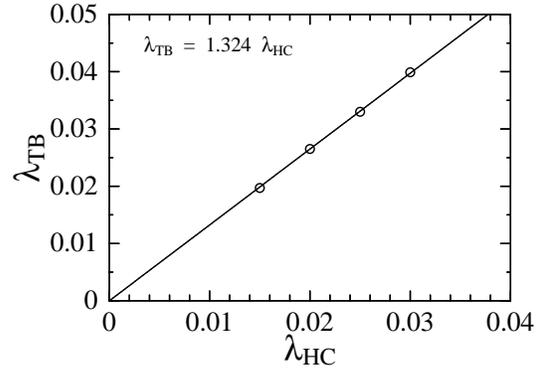}
\end{center}
\caption{ 
Linear dependence between  $\lambda_{\rm TB}$ and $\lambda_{\rm HC}$.}
\label{linear_hc_tb} 
\end{figure}

It is possible to interpret 
the relation $\Delta x \sqrt{\lambda} \sim const.$ as the quantum mechanical 
uncertainty for the excited quasiparticle under this harmonic potential.
The quasiparticle distribution in Fig.~\ref{spin_polarization} can be 
explained quantitatively by means of the tight-binding
model given in eq.~(\ref{tight_binding}). Let us consider the occupation number 
$\langle n_i^{~} \rangle = \langle c_j^{\dagger} c_j^{~} \rangle$ for the 
lowest energy one-particle state. It is possible to obtain a good approximation
of $\langle S_j^{\rm Z} \rangle$ by $\langle n_i^{~} \rangle$ if we choose
the parametrization $t = 1$ and $\mu = 2.033$.
Figure~\ref{compair_hc_tb} shows the correspondence between 
$\langle S_j^{\rm Z} \rangle$  and 
$\langle n_i^{~} \rangle$, where $\lambda_{\rm HC}^{~}$ represents the 
deformation parameter for the spin chain and $\lambda_{\rm TB}^{~}$ 
that for the tight-binding model. The value of 
$\lambda_{\rm TB}^{~}$ is determined so as to have a best fit of 
$\langle S_j^{\rm Z} \rangle$  and 
$\langle n_i^{~} \rangle$. 
Figure~\ref{linear_hc_tb} shows the relation between $\lambda_{\rm TB}^{~}$
and $\lambda_{\rm HC}^{~}$.
We can see that a simple relation 
$\lambda_{\rm TB}^{~} = 1.324 \lambda_{\rm HC}^{~}$ holds, 
where the proportional constant $1.324$ gives the correction to the
qualitative description in eq.~(\ref{one-particle_potential}).

\section{Gap Estimation by Extrapolation in $\lambda$}
\label{result2}

We have observed that the magnetic excitation of the hyperbolically deformed
$S = 1$ Heisenberg chain is bounded around the center of the system.
In this section we focus on the excitation energy 
\begin{equation}
\Delta( \lambda ) = E_1^{~}( \lambda ) - E_0^{~}( \lambda ) 
\label{gap_lambda}
\end{equation}
and investigate its dependence on $\lambda$. Here, 
$E_0^{~}( \lambda )$ is the ground-state energy and 
$E_1^{~}( \lambda )$ is the energy of the first excited state.
As it is shown in the following, $\Delta( \lambda )$ is insensitive to 
the boundary interaction parameter $J_{\rm end}$.

\subsection{Insensitivity of $\Delta( \lambda )$ with respect to $J_{\rm end}$} 

\begin{figure}[Htb]
\begin{center}
\includegraphics[width=80mm]{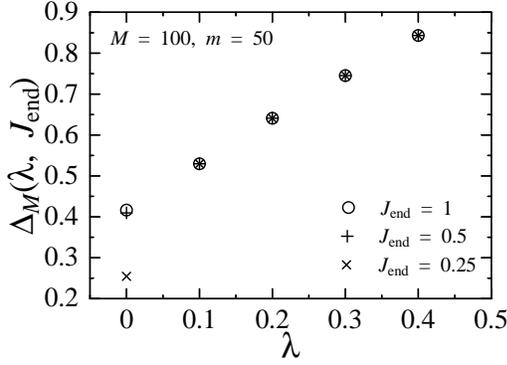}
\end{center}
\caption{
Excitation energy $\Delta_M( \lambda, J_{\rm end} )$ for the
$100$-site system when $J_{\rm end}^{~} = 0.25$, $0.5$, and $1$.}
\label{delta_lambda}
\end{figure}

Figure~\ref{delta_lambda} shows $\Delta_{M}(\lambda, J_{\rm end})$ 
for the $100$-site system, where $\Delta_{M}^{~}( \lambda, J_{\rm end}^{~} )$ is
the calculated gap for the $M$-site system when the boundary interaction is
$J_{\rm end}^{~}$.
We have chosen the set of deformation parameter 
$\lambda = 0$, $0.1$, $0.2$, $0.3$, and $0.4$, 
with $J_{\rm end}^{~} = 0.25$, $0.5$, and $1$. 
When $\lambda = 0$,  the gap $\Delta_{100}( 0, J_{\rm end} )$ is 
dependent on the value of $J_{\rm end}^{~}$. 
This is because the excited quasiparticle can reach the system 
boundary, and it is affected by the effect of $J_{\rm end}^{~}$. 
In particular, when $\lambda = 0$ and $J_{\rm end}^{~} = 0.25$ 
the quasiparticle is even localized near the system boundary. 
Appearance of this surface excitation tells that 
we need to avoid a parameter range of $\lambda\ll 0.1$ 
with $J_{\rm end}^{~}=0.25$. 

On the other hand when $\lambda \ge 0.1$, 
the excited quasiparticle cannot reach the system boundary 
as shown in Fig.~\ref{spin_polarization}, and 
the effect of $J_{\rm end}^{~}$ on $\Delta_{100}( \lambda, J_{\rm end} )$ is negligible 
for this excitation. 
In this way, the hyperbolic deformation has an effect of separating  
elementary excitations from the system boundary. 

\begin{figure}[Htb]
\begin{center}
\includegraphics[width=90mm]{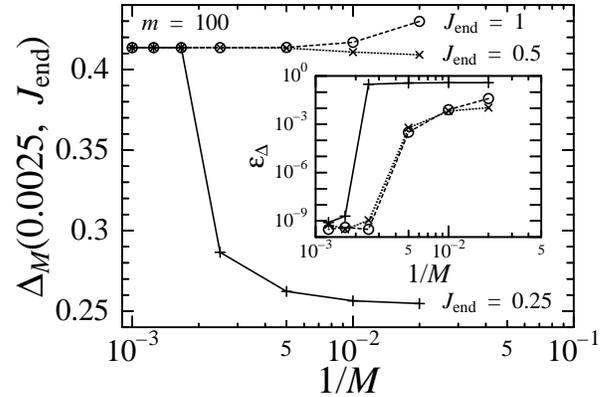}
\end{center}
\caption{
Excitation energy $\Delta_M( 0.0025, J_{\rm end} )$ as a function of 
$1/M=1/(2N)$ under the condition $J_{\rm end} =$ 0.25, 0.5, and 1. 
The inset shows $\epsilon_\Delta$ in eq.~(\ref{conv_err}). }
\label{gap_jrj}
\end{figure}

To fix a desirable parameter range of $M$ and $\lambda$, 
in order to decouple quasiparticle from the boundary, 
let us observe the system size dependence of $\Delta_M( \lambda, J_{\rm end} )$ for the case 
$\lambda = 0.0025$, which is the smallest one used in the following analysis. 
The system size is increased up to $M=2N = 1000$. 
We introduce a quantity judging an error 
\begin{equation}
\epsilon_\Delta^{~} = 
\left| \frac{\Delta_{M}^{~}(\lambda, J_{\rm end}^{~} )^{~}_{~}}{
\Delta_{1000}^{~}(\lambda, 0.25 )} - 1 \right |
\label{conv_err}
\end{equation}
In the parameter region 
where $\epsilon_\Delta^{~}$ is close to zero,
we can say that the boundary effect is removed. Figure~\ref{gap_jrj} 
shows $\Delta_M(0.0025, J_{\rm end})$ with respect to $1/M=1/2N$ 
at $J_{\rm}=0.25$, $0.5$ and $1$. 
When $J_{\rm end}^{~} = 0.25$ and 
when the system size is small, 
the boundary excitation is detected 
as appearance of smaller gap around $\Delta_{M}(0.0025, 0.25) \simeq 0.26$. 
Because the strength of the boundary interaction is 
$J_{\rm end}^{~} \, \cosh \lambda N$, the energy of the boundary excitation 
increases with the system size. 
Finally the bulk excitation $\Delta(\lambda = 0.0025)$ is detected 
as a gap of $\Delta_M(0.0025, J_{\rm end})\simeq 0.41$ 
in the neighborhood of $M=2N = 1000$,
From $\epsilon_\Delta^{~}$ shown in the inset, we can say that the effect
of $J_{\rm end}^{~}$ to $\Delta_M( \lambda, J_{\rm end} )$ is less than $10^{-9}$ 
when the system size reaches 1000. Note that $\epsilon_\Delta^{~}$ is
greater than $10^{-10}$ since we set the convergence threshold of the
eigensolver, 
which is used in the finite-size sweeping process, to the value $10^{-8}$. 
When it is necessary, we decrease the threshold down to $10^{-10}$ in the
following numerical calculations.

\subsection{$\lambda$ dependence of the energy gap}

We have erased the effect of system boundary from the elementary
excitation by the hyperbolic deformation. Thus the estimation process 
for the Haldane gap $\Delta$ is reduced to the extrapolation of 
$\Delta( \lambda )$ with respect to the deformation parameter $\lambda$.
As we will see, 
the difference $\Delta( \lambda ) - \Delta( 0 )$ 
is nearly proportional to $\lambda$,
where the dependence is consistent with the picture of 
the shallow bound state appearing in \S 3. 

In the following analysis, we use $\Delta_{600}( \lambda, 1 )$ 
as the bulk excitation $\Delta(\lambda)$, 
because the estimation value $\Delta_M( \lambda, 1 )$
is not changed within numerical precision we require
when the system size $M$ runs from 600 to 1000.
Fitting 2nd-order polynomials to the shown data, 
we obtain $\Delta(0) = 0.41047941(1)$ when $m = 160$ and
$\Delta(0) = 0.41047931(1)$ when $m = 180$, where
the numbers in the parenthesis represent the fitting error.
As we have discussed, we may choose
the smaller one $\Delta(0) = 0.41047931$ 
as the estimated upper bound of the Haldane gap $\Delta$. 

\begin{figure}[Htb]
\begin{center}
\includegraphics[width=80mm]{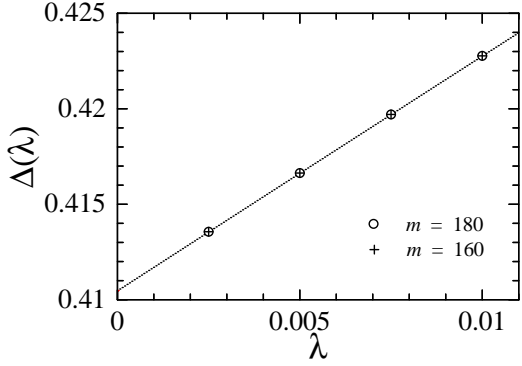}
\end{center}
\caption{Excitation energy $\Delta( \lambda )$ when $J_{\rm end}^{~} = 1$. 
The dotted line shows the result of fitting to the second-order polynomial.}
\label{poly_delta} 
\end{figure}

\begin{figure}[Htb]
\begin{center}
\includegraphics[width=80mm]{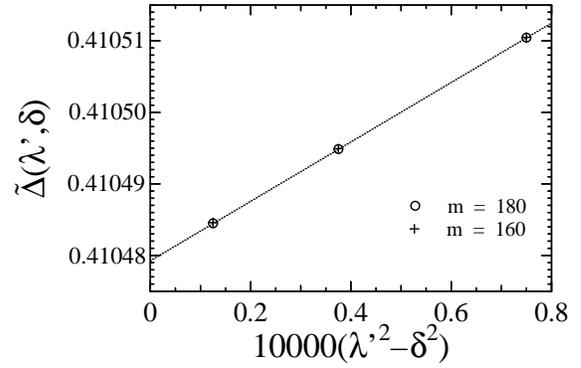}
\end{center}
\caption{
${\tilde \Delta}( \lambda', \delta )$ in eq.~(\ref{conv_acc}) 
with respect to $\lambda'$.
The dotted line represents the linear fitting.}
\label{delta_tilde} 
\end{figure}

Let us check the precision of the value $\Delta(0) = 0.41047931(1)$ 
that we have obtained from the independent data analysis.
Let us assume that  $\Delta( \lambda )$ can be represented in
terms of a polynomial in $\lambda$
\begin{equation}
\Delta(\lambda) = \Delta( 0 ) + a\lambda + b\lambda^2 + O(\lambda^3) \, .
\label{poly_gap}
\end{equation}
A way of estimating $\Delta( 0 )$ efficiently is to consider 
the derivative between two values of deformation parameters
\begin{equation}
 \frac{ \lambda_2^{~} \Delta\left( \lambda_1^{~} \right) -
  \lambda_1^{~} \Delta\left( \lambda_2^{~} \right)}{ \lambda_2^{~} - \lambda_1^{~} } 
= \Delta( 0 ) -b \lambda_1^{~} \lambda_2^{~} + O\left( \lambda^3_{~} \right) \, ,
\label{derivative}
\end{equation}
which does not contain first order term in $\lambda$. 
Introducing the notation
 $\lambda' = \left( \lambda_2^{~} + \lambda_1^{~} \right) / 2$ and 
 $\delta = \left( \lambda_2^{~} - \lambda_1^{~} \right) / 2$, this derivative can be
written as
\begin{equation}
\Delta( 0 ) - b \left( {\lambda'}^2_{~} - \delta^2_{~} \right) 
+  O\left( {\lambda'}^3_{~} \right)  = 
{\tilde \Delta}( \lambda', \delta ) \, .
\label{conv_acc}
\end{equation}
Using the data shown in Fig.~\ref{poly_delta}, we calculate ${\tilde \Delta}( \lambda', \delta )$ 
for neighboring $\lambda$s 
and plot the result  in Fig.~\ref{delta_tilde}. It is obvious that the third order 
correction $O( {\lambda'}^3_{~} )$ is very small in the shown parameter
area, and  ${\tilde \Delta}( \lambda', \delta )$ is almost linear in
${\lambda'}^2_{~} - \delta^2_{~}$. By use of linear extrapolation 
we obtain ${\Delta}( 0 ) = 0.41047940(2)$ when $m = 160$ and
${\Delta}( 0 ) = 0.41047931(1)$ when $m = 180$, where we have
shown the fitting error in the parenthesis. 
As we have done for the previous estimation for the upper bound, 
considering $m$-dependence of the value, 
we choose ${\Delta}( 0 ) = 0.41047931$ as the candidate. 
This value is the same as the number obtained in the previous paragraph. 


\section{Gap Estimation by the Sequence Interval Squeeze Method}

In this section, we present another new result 
on upper and lower bounds of the true Haldane gap 
using an examination of numerical-diagonalization data. 
The reason for the usage of the numerical-diagonalization 
technique is to keep the high precision in the original 
numerical data. 
Having an independent estimate by a complementary approach 
to the DMRG calculation, we obtain definite values 
on two bounds, which are concluded in the last section. 

Quite recently Nakano, one of the authors, and Terai proposed 
a new way to create an increasing (decreasing) sequence from 
monotonically decreasing (increasing) sequences 
of numerical data, which was immediately applied 
to estimation of the Haldane gap 
with high accuracy.~\cite{Nakano} 
A noticeable superiority of their analysis is that 
the method enables one to estimate a lower (upper) bound 
simultaneously with an upper (lower) bound of the gap. 

Here, we review this estimation method. 
First, we consider the initial sequence $A^{(0)}_{M}$ 
for $M = 2, 4, 6, \dots$, 
which is convergent to $A^{(0)}_{\infty}$. 
Suppose that we can generate   
$A^{(k)}_{M}$ for $k = 1, 2, 3, \dots$ from 
$A^{(0)}_{M}$ by some method 
so that the new sequences $A^{(k)}_{M}$ have the same 
limit, namely $A^{(k)}_{\infty}=A^{(0)}_{\infty}$ for all $k$. 
Supposed also that the set of the sequences $A^{(k)}_{M}$ has 
the following properties, 
\begin{itemize}
\item[(i)] $A^{(k)}_{M}$ is monotonic with respect to $M$. 
\item[(ii)] $\xi^{(k)}_{M}$ increases with $M$; 
    $\xi^{(k)}_{M+2} > \xi^{(k)}_{M}$, 
\item[(iii)] $\xi^{(k)}_{M}$ decreases with $k$; 
    $\xi^{(k+1)}_{M} < \xi^{(k)}_{M}$, 
\end{itemize}
where $\xi$ is the decay length\cite{Golinelli2} given by 
\begin{equation} 
\xi^{(k)}_{M} = 2/\log 
\left(
\frac{A^{(k)}_{M-4}-A^{(k)}_{M-2}}{A^{(k)}_{M-2}-A^{(k)}_{M}}
\right) \, .
\label{xi}
\end{equation}
Nakano and Terai introduced another sequence~\cite{Nakano} 
obtained from the sequences $A^{(k^{\prime})}_{M}$ and 
$A^{(k)}_{M}$ 
which are convergent to $A^{(0)}_{\infty}$ from the same side. 
The new sequence $B^{(k)}_{M}$ is given by 
\begin{equation}
B^{(k)}_{M+1} =
\frac{A^{(k)}_{M} A^{(k^{\prime})}_{M+2} 
    - A^{(k)}_{M+2} A^{(k^{\prime})}_{M}}{
 A^{(k^{\prime})}_{M+2} - A^{(k^{\prime})}_{M}
- A^{(k)}_{M+2} + A^{(k)}_{M}} , 
\label{mod_NT}
\end{equation}
for $k>k^{\prime}$. 
The most important property of $B^{(k)}_{M}$ is that 
the new sequence is convergent to the same limit 
$A^{(0)}_{\infty}$ from the {\it opposite} side. 
(See appendix in ref.~[\ref{Nakano}].) 
Thus, there is a relation 
$B^{(k_1)}_{M_1}<A^{(0)}_{\infty}<A^{(k_2)}_{M_2}$ 
(or $A^{(k_1)}_{M_1}<A^{(0)}_{\infty}<B^{(k_2)}_{M_2}$) and 
we obtain a reliable interval 
including the limit $A^{(0)}_{\infty}$, 
which we would like to know. 
Here $k_1$, $k_2$, $M_1$ and $M_2$ are integers.
When the sequences $A^{(k)}_{M}$ and $A^{(k^\prime)}_{M}$ are 
monotonically decreasing,  
$\min_{k, M}(A^{(k)}_{M})$ and $\max_{k, M}(B^{(k)}_{M})$ 
are an upper bound and a lower one for $A^{(0)}_{\infty}$, 
respectively. 
When the direction of $A^{(k)}_{M}$ and $A^{(k^\prime)}_{M}$ 
is opposite, 
$\max_{k, M}(A^{(k)}_{M})$ and $\min_{k, M}(B^{(k)}_{M})$ 
give a lower bound and an upper one, respectively. 

An appropriate set of the sequences $A^{(k)}_{M}$ can be 
generated systematically from $A^{(0)}_{M}$ 
to make the interval narrower 
by using convergence-acceleration techniques. 
As such a technique, 
we discuss the $\varepsilon$-algorithm~\cite{Wynn,Golinelli2} 
and its generalization. 
The $\varepsilon$-algorithm provides us with 
a new sequence of one-level higher, 
by the relation between the neighboring three levels 
\begin{eqnarray}
& & 
\hspace{-5mm}
\frac{1}{A^{(k+1)}_{M} - A^{(k)}_{M-2}}
\label{mod_wynn_epsilon}
\\
&=&
\hspace{-3mm}
\frac{1}{A^{(k)}_{M-4} - A^{(k)}_{M-2}}
+\frac{1}{A^{(k)}_{M} - A^{(k)}_{M-2}}
-\frac{\alpha
}{A^{(k-1)}_{M-4}-A^{(k)}_{M-2}} ,
\nonumber 
\end{eqnarray}
for $\alpha=1$. 
This algorithm for $\alpha=1$ was 
originally developed by Wynn~\cite{Wynn}. 
In order to create the first level sequence $A^{(1)}_M$, 
we can prepare a dummy sequence $A^{(-1)}_M$ of the level $-1$, 
where all of the elements are infinitely large, 
namely $A^{(-1)}_M = \infty$. 

To know whether the acceleration is successful or not, 
the monitoring of the decay length given by eq.~(\ref{xi}) 
is important. 
One can consider that the convergence of a transformed 
sequence $A^{(k)}_M$ for $k \ge 1$ is successfully accelerated 
when all the above conditions (i)-(iii) hold. 

We put the excitation gap of the 
$M$-site $S = 1$ chain 
$\mathcal{H} = \sum_{i = 1}^{M} {\bf S}_i \cdot {\bf S}_{i+1}$
under a certain boundary condition into 
the initial sequence $A^{(0)}_M$. 
Note that the direction of the monotonic behavior 
of the sequence depends on the boundary condition adopted. 
If we impose the periodic (${\bf S}_{M+1} = {\bf S}_1$) or 
the twisted ($S^{x}_{M+1} = -S^x_1, S^{y}_{M+1} = -S^y_1, S^{z}_{M+1} = S^z_1$) boundary condition, 
the direction is different with each other.\cite{Nakano} 
What we would like to do is 
to estimate $A^{(0)}_{\infty}$ only from a finite part 
of the initial sequence. 

For long, only the systems 
under the periodic boundary condition were examined 
in most of the finite-size-scaling studies 
based on numerical diagonalization data. 
Finite-size gap of such systems usually decreases monotonically, 
when the system size is increased. 
If we apply the $\varepsilon$-algorithm 
to the monotonically decreasing sequence, 
we obtain only upper bounds of the Haldane gap. 
Nakano and Terai found that 
the excitation gap of the finite size systems 
under the twisted boundary condition 
is monotonically increasing.~\cite{Nakano} 
Thus, we should examine which is better among 
both of the boundary conditions, periodic and twisted. 

We can interpret that Nakano and Terai found 
a quite systematic approach 
to overcome limitation of the usage of 
the original $\varepsilon$-algorithm.~\cite{Nakano} 
Fundamental steps are summarized as usage of 
1) examination of different boundary conditions, 
2) acceleration of the monotonic sequences $A^{(k)}_{M}$, 
and 
3) the above new sequence $B^{(k)}_{M}$. 
If an appropriate acceleration transformation is chosen, 
one easily obtains a reliable interval including the limit 
$A^{(0)}_{\infty}$, which we would like to know. 
The interval gets narrower as the number of initial data 
is increased. 
Hereafter, we call the above procedure 
the sequence interval squeeze (SIS) method. 
Getting the bounds of both sides, 
we can quantitatively discuss the precision 
of the estimates of the Haldane gap that were reported so far. 

Nakano and Terai substituted 
the excitation gap of the finite size systems 
under the twisted boundary condition for $A^{(0)}_{M}$; 
the initial sequence is monotonically increasing. From 
the numerical data collected up to $A^{(0)}_{M=24}$ 
with the twisted boundary condition, 
the SIS method provides us 
with the estimate $\Delta = 0.4104789(13)$ 
as the Haldane gap from the acceleration process 
under $\alpha = 1$. 

Let us recall that an available acceleration transformation 
is not limited to the $\varepsilon$-algorithm with $\alpha=1$. 
Actually, the first step of $\varepsilon$-algorithm 
is equivalent to the Aitken-Shanks transformation,\cite{Shanks} 
which corresponds to $\alpha=0$. 
If one takes $\alpha=0$ for the second step and the later,  
the transformation (\ref{mod_wynn_epsilon}) is reduced to 
just an iteration of the Aitken-Shanks transformation. 
Other choices of $\alpha$ 
for the second step and the later correspond 
to different convergence acceleration transformations, 
on each of which the degree of acceleration depends. 
Thus we may optimize the convergence acceleration 
in the step 2) in the SIS method. 
If we adjust $\alpha$ within a range 
in which the above three conditions are certified, 
the strength of the acceleration can be optimized. 
In the present work, we apply $\alpha = 0.4$ 
in eq.~(\ref{mod_wynn_epsilon}) giving $A^{(k)}_{M}$ 
and $k^{\prime}=k-2$ in eq.~(\ref{mod_NT}) giving $B^{(k)}_{M}$ 
to the same numerical data with the twisted boundary condition. 
The result successfully gives an inside interval narrower 
than the result obtained by $\alpha = 1$. 

\begin{table}[Htb]
\caption{Sequence of finite-size gaps 
by eq.~(\ref{mod_wynn_epsilon}) 
with $\alpha = 0.4$.}
\label{gap_s1_tbc_a0_4}
\begin{tabular}{r|cc|cc}
\hline
$M$ & $A^{(2)}_M$ & $\xi^{(2)}_M$ & $A^{(3)}_M$ & $\xi^{(3)}_M$ \\
\hline
12 & 0.409599020 &  &   & \\
14 & 0.410158700 &  &   & \\
16 & 0.410354714 & 1.91 & 0.410442366 & \\
18 & 0.410427448 & 2.02 & 0.410465180 & \\
20 & 0.410456146 & 2.15 & 0.410473181 & 1.91 \\
22 & 0.410468261 & 2.32 & 0.410476497 & 2.27 \\
24 & 0.410473733 & 2.52 & 0.410477982 & 2.49 \\
\hline
\hline
$M$ & $A^{(4)}_M$ & $\xi^{(4)}_M$ & $A^{(5)}_M$ & \\
\hline
20 & 0.410476976 & &  \\
22 & 0.410478577 & &  \\
24 & 0.410479051 & 1.64 & 0.410479218 \\
\end{tabular}
\end{table}

\begin{table}[Htb]
\caption{Antimonotonic sequence $B^{(k)}_{M}$ 
for $\alpha=0.4$.}
\label{upperbounds_s1}
\begin{tabular}{r|c|c|c}
\hline
$M$ 
& $B^{(2)}_{M}$ 
& $B^{(3)}_{M}$ 
& $B^{(4)}_{M}$ 
\\
\hline
13 & 0.411197171 &  &  \\  
15 & 0.410724641 &  &  \\  
17 & 0.410568330 & 0.410491523 &  \\  
19 & 0.410513257 & 0.410482576 &  \\  
21 & 0.410492985 & 0.410480499 & 0.410480148 \\  
23 & 0.410485151 & 0.410479829 & 0.410479554 \\  
\hline
\end{tabular}
\end{table}

\begin{figure}[Htb]
\begin{center}
\includegraphics[width=8cm]{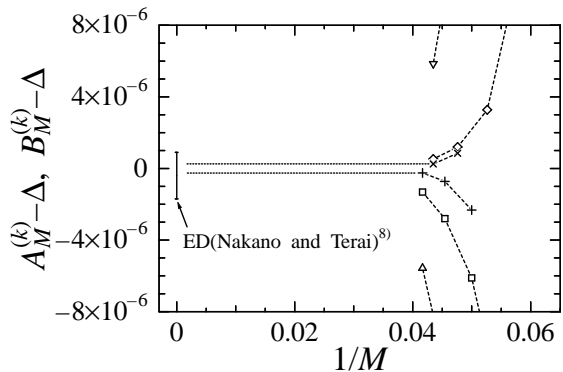}
\end{center}
\caption{Convergence of the data shown in Table \ref{gap_s1_tbc_a0_4} and 
\ref{upperbounds_s1} with respect to $M$. 
We use the common shift $\Delta=0.4104793$ for display. 
Triangles, squares, and pluses denote 
$A^{(2)}_M$, $A^{(3)}_M$, and $A^{(4)}_M$, respectively.
Reversed triangles, diamonds, and crosses denote 
$B^{(2)}_M$, $B^{(3)}_M$, and $B^{(4)}_M$, respectively. }
\label{fig14}
\end{figure}

Let us list all the $A^{(k)}_M$ created 
by eq.~(\ref{mod_wynn_epsilon}) in Table II.
Sequences $A^{(0)}_M$ and $A^{(1)}_M$ are not presented 
because they do not change due to a variance of $\alpha$. 
As it is observed, 
$A^{(k)}_M$ and $\xi^{(k)}_M$ satisfy 
the set of the three conditions 
for the successful convergence acceleration. 
We then create another sequence 
$B^{(k)}_{M}$ by eq.~(\ref{mod_NT}) and show them in Table III.
One can clearly observe that $B^{(k)}_{M}$ is monotonically 
convergent from the side opposite to $A^{(k)}_{M}$. 
Figure~\ref{fig14} depicts the $1/M$ dependence 
of these sequences. 
The present result gives 
a new interval $[0.410479051,~0.410479554]$ 
for an estimate for the Haldane gap $\Delta$, 
which is narrower than the reported one in ref.~[\ref{Nakano}].

\section{Conclusions and Discussions}

We have observed the magnetic excitation of the $S = 1$ Heisenberg  
chain,
whose exchange coupling is deformed hyperbolically. The magnetic  
quasiparticle
is weakly bounded in the neighborhood of the center of the system.
It is shown that the excitation energy $\Delta( \lambda )$ is  
nearly linear in $\lambda$,
and the extrapolation to $\lambda = 0$ gives an estimate of the  
Haldane gap,
as precise as that obtained by the SIS method.
Since the quasiparticle does not reach the system boundary under the
hyperbolic deformation, one does not have to pay special attention to  
the
boundary interaction strength.

\begin{figure}[Htb]
\begin{center}
\includegraphics[width=80mm]{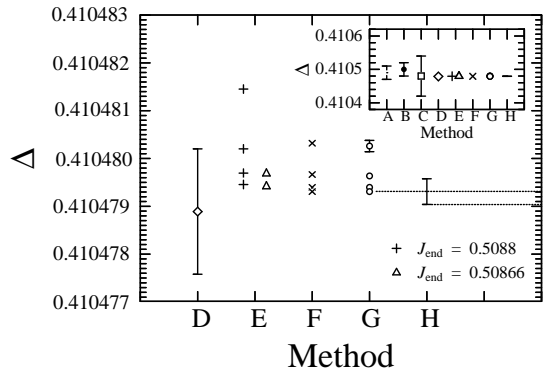}
\end{center}
\caption{
The values of the Haldane Gap which have been reported
and that calculated in this article.
A scale of the perpendicular axis is adjusted
so that all error bar are shown in the inset.
}
\label{gap_method}
\end{figure}

Here, let us summarize the estimated values of the Haldane gap we have  
obtained
with those reported so far.
Figure~\ref{gap_method} shows ${\Delta}( 0 )$ obtained by
\begin{itemize}
\item[A.] Lanczos method + convergence acceleration in ref.~[\ref{Lan}]
\item[B.] DMRG applied to undeformed system in ref.~[\ref{Huse}]
\item[C.] Monte Carlo simulation in ref.~[\ref{MC}]
\item[D.] Lanczos method + SIS in ref.~[\ref{Nakano}]
\item[E.] DMRG applied to undeformed system in \S 3.
\item[F.] Deformation analysis in \S 4 with polynomial fitting.
\item[G.] Deformation analysis in \S 4 with derivatives.
\item[H.] Lanczos method + SIS in \S 5.
\end{itemize}
Each method among E at $J_{\rm end}=0.5088$, F, G
has four data points when $m = $120, 140, 160 and 180.
In addition, there are two data points under the conditions
$m = 160$ and 180 for the method E at $J_{\rm end}=0.50866$.

In Fig.~\ref{gap_method}, 
$\Delta(0)$ given by one of methods E-G always monotonically 
decreases, when $m$ is increased.
We know that the energy lift by the cut-off effect in the excited state is
larger than that in the ground state. Besides, 
the systems treated in \S 4 have enough sizes, where 
any boundary effects are eliminated. 
Thus, we may suppose that 
the estimated values of $\Delta (0)$ 
converge monotonically with respect to $m\ge 120$. 
Note that the difference between
$\Delta(0) = 0.41047941(1)$ when $m = 160$ and
$\Delta(0) = 0.41047931(1)$ when $m = 180$ is big enough
compared with the fitting error. 

It should be noted that the SIS estimation~\cite{Nakano}
gives both the lower and the upper bound for the Haldane gap.
The values given by the item H provide us the best bounds
with this method. 
When we choose $m = 180$, our estimations in E-G 
are always inside the interval given by the item H. 

Thus, we may safely conclude that
$0.41047931$ should be a better upper bound of the Haldane gap 
than the value given by the item H.
Following this discussion, we conclude that the Haldane gap
is in $[0.41047905,~0.41047931]$.
Looking at Figure~\ref{gap_method},
we may also suppose that the 
actual Haldane gap is closer to the upper bound
than the lower bound.

For the hyperbolic deformation, the choice of $m$
is the remaining single parameter determining the accuracy. 
If we consider the $m$ dependence of the gap, we 
may also construct a sequence interval squeeze technique
by generalizing the method given in \S 5. 
We can thus find 
that the use of the hyperbolic deformation is
one of the efficient tool to detect the excitation gaps of one-dimensional
quantum systems.

One of future subjects is numerical gap estimation of
$S=2$ antiferromagnetic Heisenberg chain.
Let us consider appropriate boundary conditions based on
the valence bond solid (VBS) picture for this case.
A possible simple choice is to put $S = 1$ spins at the both ends of the  
system.
Another simple choice is to put two $S=1/2$ spins at the each end of the  
system, 
where the bond configuration is the form of the letter Y.
In addition, a slightly complex choice is
to reduce the length of spin by amount of
1/2 site by site, i.e. to put $S=3/2$, $S=1$, and $S=1/2$ spins at the  
boundary.
For each candidate of the boundary spin arrangement,
one has to consider the parametrization of the bond strength.
In this way, the tuning of the boundary condition for the
$S=2$ chain is more complicated than that of $S=1$ chain.
Therefore, an efficient arrangement of additional boundary spins
has not been reported.
When the system is deformed hyperbolically, however,
the problem of parameterization could be put in a extrapolation of $ 
\lambda$.

Another one is to find out classical analogue of the hyperbolic  
deformation
for 2D statistical models. A candidate is the hyperbolic lattice  
models studied
so far,~\cite{Gendiar1,Gendiar2,Gendiar3,Gendiar4} but in those  
models
only discrete values of $\lambda$ are allowed.
To construct a class of models that has
appropriate structure for the DMRG applied to classical systems~ 
\cite{Nishino2}
would be important for the further study.

\acknowledgements

The authors thank to Okunishi, Gendiar, and Krcmar for valuable discussions.
This work was partly supported by Grants-in-Aid 
from the Ministry of Education, Culture, Sports, Science 
and Technology (MEXT) (No.~19540403 and No.~20340096), and 
the Global COE Program (Core Research and Engineering of Advanced 
Materials-Interdisciplinary Education Center for Materials Science), 
MEXT, Japan. 
A part of the computations was performed using facilities 
of the Information Initiative Center, Hokkaido University 
and the Supercomputer Center, 
Institute for Solid State Physics, University of Tokyo.


\begin{thebibliography}{99}
\bibitem{diverge} The distance $( \cosh \nu x ) \, \Delta \tau$ diverges in the 
limit $| x | \rightarrow \infty$. This divergence can be controlled by considering
finite width $R$ and keep $| x | < R$.
\bibitem{long} This kind of difference cannot be neglected when one 
considers long-range interactions in the framework of the hyperbolic deformation.
\bibitem{commute} Note that the first and the second terms in the r.h.s. of eq.~(\ref{tight_binding}) do
not commute.
\end{thebibliography}
\end{document}